\documentclass[manuscript]{acmart} 

\usepackage{enumitem}
\usepackage{amsmath}
\usepackage{multirow}
\usepackage{xcolor}

\newcommand\cnot[1]{%
  \mathrel{\ooalign{\hfil$#1$\hfil\cr\hfil$/$\hfil\cr}}}

\AtBeginDocument{%
  \providecommand\BibTeX{{%
    \normalfont B\kern-0.5em{\scshape i\kern-0.25em b}\kern-0.8em\TeX}}}

\copyrightyear{2022} 
\acmYear{2022} 
\setcopyright{rightsretained} 
\acmConference[IUI '22]{27th International Conference on Intelligent User Interfaces}{March 22--25, 2022}{Helsinki, Finland}
\acmBooktitle{27th International Conference on Intelligent User Interfaces (IUI '22), March 22--25, 2022, Helsinki, Finland}\acmDOI{10.1145/3490099.3511137}
\acmISBN{978-1-4503-9144-3/22/03}

\begin{document}

\title[Emblaze: Illuminating Machine Learning Representations]{Emblaze: Illuminating Machine Learning Representations through Interactive Comparison of Embedding Spaces}

\author{Venkatesh Sivaraman}
\email{venkats@cmu.edu}
\orcid{1234-5678-9012}

\author{Yiwei Wu}
\email{yiweiw@andrew.cmu.edu}

\author{Adam Perer}
\email{adamperer@cmu.edu}
\affiliation{%
  \institution{Carnegie Mellon University}
  \streetaddress{5000 Forbes Ave}
  \city{Pittsburgh}
  \state{Pennsylvania}
  \country{USA}
  \postcode{15213}
}


\begin{abstract}
    Modern machine learning techniques commonly rely on complex, high-dimensional embedding representations to capture underlying structure in the data and improve performance. In order to characterize model flaws and choose a desirable representation, model builders often need to compare across multiple embedding spaces, a challenging analytical task supported by few existing tools. We first interviewed nine embedding experts in a variety of fields to characterize the diverse challenges they face and techniques they use when analyzing embedding spaces. Informed by these perspectives, we developed a novel system called Emblaze that integrates embedding space comparison within a computational notebook environment. Emblaze uses an animated, interactive scatter plot with a novel Star Trail augmentation to enable visual comparison. It also employs novel neighborhood analysis and clustering procedures to dynamically suggest groups of points with interesting changes between spaces. Through a series of case studies with ML experts, we demonstrate how interactive comparison with Emblaze can help gain new insights into embedding space structure.
\end{abstract}



\begin{teaserfigure}
\includegraphics[width=\textwidth]{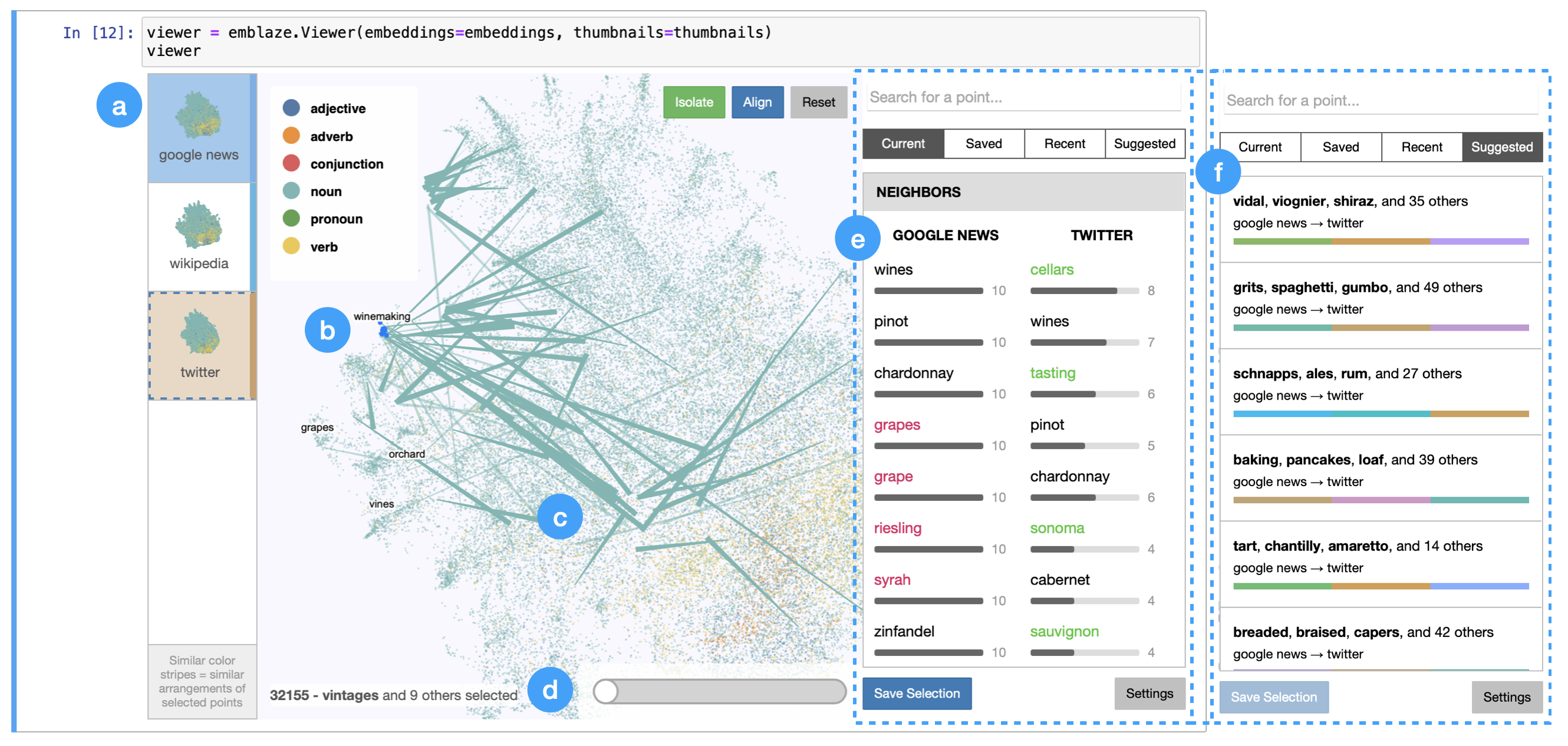}
\caption{Interactive comparison of embeddings for 50,000 words from three different corpora (a), using Emblaze within a Jupyter notebook. Given a selection, in this case a group of words about winemaking (b), the Star Trail visualization (c) highlights points near the group in the high-dimensional space whose neighborhoods change significantly. The plot can be manually interpolated between the two embedding spaces using a slider (d). In the sidebar, the neighborhood comparisons for the current selection (e) show a greater emphasis on recreational wine-related activities in Twitter data, such as ``tastings'' and ``Sonoma.'' The sidebar can alternatively display Suggested Selections relevant to the current visualization state (f), including clusters of wine varieties and other beverages.}
\label{fig:teaser}
\end{teaserfigure}

\maketitle

\section{Introduction}

Most state-of-the-art machine learning (ML) techniques work by learning an expressive representation of their input data in a high-dimensional vector space. This representation, also known as an embedding, reflects both the structure of the dataset and the task used to train the model. Embedding representations have increasingly been used to improve performance on a plethora of tasks thanks to deep neural networks such as transformers, which leverage vast quantities of often-unlabeled data to learn highly nuanced representations. However, embedding spaces can acquire unpredictable and undesirable structural features during training, learning shortcuts or biases in the data to perform better on the learning task \cite{Bolukbasi2016,Steed2021}. Model builders and data scientists need effective tools to probe the structure of their embeddings and to help them choose the best representation for their task.

Although a variety of tools have been developed to visualize and probe embeddings, the problem of extending these techniques to compare across \textit{multiple} embedding spaces remains an open challenge. For instance, many popular embedding analysis tools use dimensionality reduction (DR) techniques such as tSNE and UMAP to generate 2D scatter plots of the embedding space \cite{Smilkov2016,Ovchinnikova2019,Li2018}. These visualizations are well-suited to give a high-level overview of an individual embedding space, particularly when combined with augmentations that highlight distortions due to DR \cite{Nonato2019}. However, visualizations that juxtapose two DR visualizations side-by-side (e.g. \cite{Cutura2020,Boggust2019,Arendt2020}) may become visually taxing and confusing when there are thousands of points displayed and different degrees of distortion in each plot. This juxtaposition approach is also typically limited to two embedding spaces to avoid overwhelming the user, but comparing more than two spaces may be necessary for some model comparison tasks (such as diachronic word embeddings \cite{Hamilton2016}).

To mitigate the complexity of comparing embedding spaces at a global scale using DR, comparison tools such as the Embedding Comparator \cite{Boggust2019} and embComp \cite{Heimerl2020} have incorporated more focused visualizations that enable comparisons on small neighborhoods of points. These features do address the need for users to analyze local differences between embedding spaces, but the task of \textit{finding} relevant neighborhoods to compare remains challenging. In the absence of pre-formed hypotheses about where in the embedding space to look, a tool that can adaptively guide analysts to interesting comparisons may be necessary.

In this work, we first present findings from interviews with nine embedding space experts from domains such as language modeling, computational biology, and dimensionality reduction. These interviews showed that model builders tend to rely on \textit{ad hoc} workflows to analyze embedding spaces, partially because many are skeptical of DR's ability to support their analytical needs. In particular, experts desire tools to build a much richer understanding of their embedding spaces, moving beyond the fixed distance metrics often assumed in DR towards much more complex and nuanced notions of similarity and uncertainty. These observations, combined with the areas of opportunity we identified from prior embedding comparison systems, led us to develop a visual embedding analysis tool that unifies common needs for comparison throughout the processes of model building and dimensionality reduction (see Table \ref{tab:compare_levels}).

\begin{table*}[]
\begin{tabular}{lp{5cm}p{7cm}}
\toprule
               & \textbf{Comparison Target}                             & \textbf{Example Research Question}                                                                               \\
\midrule
Representation & Input data                                    & What differences in learned entity relationships would arise from training on different datasets?        \\
               & Pre-trained models                            & Which pre-trained model best captures the structure in my domain-specific dataset? \\
               & Fine-tuning procedure                         & How does fine-tuning affect the embedding space of my pre-trained model?                                \\
               & Model architecture                            & Which type of model results in the most sensible clustering of my validation dataset?                   \\
               & Model hyperparameters                         & How many parameters does my model need to adequately represent the structure in my data?                \\
               & Intermediate layer representations                            & How do the neural network's activation patterns for the training data differ from layer to layer?                   \\
               & Random effects from model initialization      & What patterns in my embedding space are reliable, and which ones are likely due to noise?               \\
\midrule
Projection     & Dimensionality reduction (DR) technique       & Which DR technique produces a more expressive depiction of my data?                                     \\
               & DR parameters           & Which value of the tSNE perplexity parameter creates the best visual clustering for my data?             \\
               & Random effects from DR initialization & How consistent are the neighborhoods and clusters created by this DR algorithm? \\
\bottomrule
\end{tabular}
\caption{Use cases for embedding space comparison that arise during ML model development and visualization, generated from formative interviews, case studies, and practical experience using Emblaze. In each of these comparison tasks, model builders seek to understand the effect of the comparison target on the overall structure and relationships encoded in their embedding spaces. Emblaze supports comparison for all of these tasks through a unified set of visualization techniques.}
\label{tab:compare_levels}
\end{table*}

The resulting system, which we call Emblaze, is a Python framework and web-based visual interface that can be run within a computational notebook environment, making it easy to import and visualize heterogeneous data in tandem with \textit{ad hoc} workflows. Emblaze is centered around an animated, interactive DR scatter plot, which can be filtered and aligned in place to support navigation of different regions of the space. It incorporates novel visual augmentations that summarize changes for both pairs and larger sets of embedding spaces, and it dynamically suggests clusters of points that exhibit interesting changes. Through case studies with ML experts, we demonstrate the utility of these capabilities not just for comparing models, but also for understanding individual model behavior and reasoning about the effects of 2D projection. 

Concretely, the contributions of this work are the following:

\begin{enumerate}
    \item A series of semi-structured interviews with nine embedding experts in a variety of fields, including natural language processing, computational social science, computer vision, and computational biology. The resulting qualitative analysis expands upon prior need-finding studies by probing experts' viewpoints on specific embedding analysis and comparison techniques (such as clustering, dimensionality reduction, and embedding space alignment), and by identifying practitioners' perspectives and challenges as pertaining to their unique fields.
    \item A comprehensive system, called Emblaze, to compare embedding spaces within a computational notebook environment. The tool improves on previous embedding comparison systems by supporting comparison across several spaces and at many stages of the model building pipeline. It also introduces new techniques to surface points and clusters with interesting changes, and facilitates rapid iteration through its lightweight notebook-based interface. Emblaze is open source and publicly available.
\end{enumerate}

\section{Background and Related Work}

In this paper, we specifically focus on the problem of \textit{embedding space comparison}, which can be defined as the comparison of multiple high-dimensional representations of the same set of $N$ objects. When designing for comparison, it is important to note that comparative tasks often demand fundamentally different approaches than more general exploratory or analytical ones \cite{Gleicher2011}. However, we hypothesize that given the complexity and opaqueness of embedding spaces, comparison may be beneficial or even necessary to help users understand \textit{individual} embedding spaces. Therefore, in the following sections we briefly review strategies for visualizing dimensionally-reduced data and embeddings alongside the relevant efforts to extend those methods to comparative tasks.



\subsection{Dimensionality reduction visualization}

All dimensionality reduction techniques attempt to project an $N\times D$ matrix of $N$ observations in $D$ dimensions to a lower-dimensional $N\times d$ representation, while attempting to preserve relationships between observations \cite{VanDerMaaten2009}.\footnote{In this paper, we use \textit{embeddings} to refer to high-dimensional vector representations of objects, and \textit{projections} for dimensionally-reduced representations of those embeddings.} Because $d$ is usually much too low to capture all of the meaningful variation in the dataset, DR techniques typically make tradeoffs on the fidelity of the projection at different distance scales. For example, PCA (principal component analysis) and classical MDS (multidimensional scaling) are considered better at capturing global structure, while tSNE (t-distributed Stochastic Neighbor Embedding) and UMAP (Uniform Manifold Approximation and Projection) prioritize nearest-neighbor structure and clusters \cite{Nguyen2019}. Even for a single DR technique, parameter choices and random initializations can often dramatically affect the result, creating a boundless array of possible projections to choose from.

Visual interaction techniques to help analysts explore and evaluate DR projections have been reviewed extensively elsewhere \cite{Nonato2019,Sacha2017}, so we provide only a brief summary of relevant approaches here. For example, visual augmentations to highlight distortions in DR projections have included superimposed heat maps \cite{Aupetit2007,Seifert2010}, lines whose lengths indicate the degree of projection error \cite{Stahnke2016}, and animations between projection axes \cite{Elmqvist2008}. In addition, a few systems have incorporated multiple projections, either for interactive parameter selection \cite{Fujiwara2021,Pagliosa2015} or to compare proximity relationships in different variants \cite{Cutura2020}. In this work, we attempt to extend some of these techniques to the problems specific to embeddings in ML, including dataset scale and the need to compare more than two spaces.

\subsection{Embedding comparison across domains}

Visual analytics work on learned embedding spaces (using DR or otherwise) has most frequently focused on textual embeddings \cite{Heimerl2018,Liu2018}, particularly in light of the hidden biases often found in word embeddings \cite{Bolukbasi2016}. Some comparison tools have been developed to analyze and improve word embedding methods \cite{Rong2018,Chen2018}, but the more common use case for word embedding comparison is in holding the modeling procedure constant and analyzing semantic differences from different corpora \cite{Hamilton2016,Kulkarni2015,Toshevska2020}. For example, an early prototype of Emblaze was used as part of a visual interface to compare semantic changes in clinical concepts related to COVID-19 \cite{covidconcepts2021}. Note that prior qualitative comparison techniques have typically only been able to visualize small, curated subsets of the embedding spaces, and do not well support comparison of groups of points.

In the life-sciences domain, embeddings are frequently used to represent gene expression levels in cells \cite{Becht2019} and human genetic profiles \cite{tsne_li2017}, among other applications. Visual comparison tools have been developed for the common task of clustering and interpreting feature values in these embeddings \cite{Kiselev2019,LYi2015}. However, the widespread use of static DR-based visualizations in computational biology has prompted criticism, including new proposed DR techniques \cite{Narayan2021,Ding2018} and calls to avoid DR-based analysis entirely \cite{Chari2021}. It is worth noting that with the notable exception of Sleepwalk, an interactive R-based tool for global-scale embedding exploration \cite{Ovchinnikova2019}, the potential for interactive visualizations that enable comparison of multiple DR projections appears to be under-explored in computational biology.

Large-scale embedding spaces for images, such as those learned by convolutional neural networks, are an important application area for which few systems have specifically been developed. The difficulty of auditing such large spaces as well as the presence of class labels may explain why embedding visualization is typically eschewed in favor of other model inspection strategies \cite{Hohman2019}; tools that do incorporate DR have primarily focused on probing image models during training \cite{Chung2016,Pezzotti2018}. However, image models can still be analyzed using general-purpose embedding visualization systems, such as the Embedding Projector \cite{Smilkov2016} and Latent Space Cartography \cite{Liu2019}.

\subsection{Visual tools for embedding comparison}

This work builds upon a small number of research systems that have been developed to support general-purpose embedding comparison. Systems by Li et al. \cite{Li2018} and Heimerl et al. \cite{Heimerl2020}, for example, utilize summary visualizations of embedding metrics, neighborhood views, and DR plots to facilitate comparison. Meanwhile, Parallel Embeddings \cite{Arendt2020} utilizes a novel clustering-based visualization to highlight correspondences between embeddings, and the Embedding Comparator \cite{Boggust2019} features PCA plots of the neighborhoods around selected points that change the most. The latter two approaches offer greater simplicity, but limit analysis to two embedding spaces at the cluster or individual-point level, respectively. 

Notably, all of these systems require the user to spend time browsing the visualizations in order to find meaningful comparisons, which may be difficult for large datasets. Moreover, when these systems do surface candidate points for comparison, the techniques used are limited to individual points and cannot easily be generalized to clusters \cite{Boggust2019}. As we discuss in our expert interviews, direct paths to interesting comparisons at varying granularities may be a key factor in gaining better insight into large unlabeled datasets.

\section{Expert Interviews}

To characterize how embedding spaces are currently analyzed and compared, we conducted semi-structured interviews with 9 embedding experts across a variety of domains. As listed in Table \ref{tab:interview_participants}, the majority of participants worked with language-based models, while others were experts in computational biology, computer vision, multimodal machine learning, and signal processing. All participants except one (P7, a DR expert) were situated within machine learning practice rather than DR, consistent with our goal of examining how experts understand embeddings beyond visually projecting them. Interviews were conducted on Zoom and lasted 53 minutes on average. Transcripts were coded and analyzed to answer three primary research questions:

\begin{enumerate}
    \item What are domain experts' overall perspectives on the role of embedding analysis in their work? (Sec. \ref{sec:interview-overall})
    \item What techniques do experts use to analyze embedding spaces individually, and what challenges are posed by those techniques? (Sec. \ref{sec:interview-single})
    \item What are experts' current approaches and needs for embedding space comparison? (Sec. \ref{sec:interview-comparison})
\end{enumerate}
Finally, we synthesized across these three sub-analyses to produce a set of themes and design goals, which we present in Sec. \ref{sec:design-goals}.


\begin{table}[]
    \centering \small
    \begin{tabular}{rll}
        \toprule Participant & Role & Domain(s) \\
        \midrule
        P1 & graduate student    & computational social science                            \\
        P2 & graduate student    & language                            \\
        P3 & industry researcher & language, computer vision           \\
        P4 & industry researcher & language                            \\
        P5 & graduate student    & computational biology          \\
        P6 & industry researcher & signal processing              \\
        P7 & graduate student    & computational biology          \\
        P8 & industry researcher & language, signal processing         \\
        P9 & professor           & computer vision, multimodal ML \\
         \bottomrule
    \end{tabular}
    \caption{Summary of roles and application domains of interview participants.}
    \label{tab:interview_participants}
\end{table}

\subsection{Overall perspectives on embedding analysis}\label{sec:interview-overall}

Most of the embedding experts we interviewed expressed a largely unmet need for tools to help them understand their models: ``the tools are very crude, and you're kind of just getting small clues from those little [ad hoc] techniques'' (P8). However, consistent with the diversity in their backgrounds, participants varied considerably in the purposes and associated levels of granularity that understanding embedding spaces entailed. 

In some cases, embedding analysis plays a distant secondary role to task-specific performance metrics; understanding the embedding space is often only useful to debug an underperforming model (P4). Participants who voiced this opinion do feel that other validation techniques are sometimes necessary, but they fall back to quantitative performance metrics because they do not have clear ways to evaluate embedding quality on very large datasets (P4, P8). However, as P8 noted, the way quantitative metrics are defined can sometimes obscure large-scale properties of the data, rendering them as ``fuzzy'' and incomplete as qualitative examples.

Other participants expressed that embedding analysis was an important, even essential aspect of their work (P1, P2, P6, P7). These experts advocated for a fairly rigorous approach to embedding analysis, in which they would define hypotheses about the structure of the embedding space and test them using a combination of existing tools and \textit{ad hoc} algorithms. However, these participants are also concerned that the methods they are currently using are too ``hand-wavy'' (P1) and that their observations may not reflect real patterns in the embedding space structure (P1, P2). Furthermore, their analytical techniques depend on the presence of previously-known points of interest, which they may not have when exploring new datasets (P7).

\subsection{Approaches and needs for single embedding space analysis}\label{sec:interview-single}

We spent considerable time probing participants' general approaches to embedding analysis because as discussed in the next section, their experience with embedding \textit{comparison} was much more limited. The techniques that these experts employ for embedding analysis therefore present useful starting points for our proposed comparison techniques.

\subsubsection{Dimensionality reduction visualization.} All participants described having used DR techniques such as tSNE and UMAP in their current or past work to assess the overall structure of an embedding space. Experts in computational biology in particular use DR extensively, not only to validate that their embedding models place similar cells close together but also to ``create the figures that biologists can appreciate'' (P5). However, several participants were skeptical of assessing the quality of a model by looking at how tightly clustered its DR plot is. For example, P3 and P8 were concerned about the tendency of nonlinear DR techniques to create spurious clustering effects, while P5 noted that a poorly-clustered DR plot does not necessarily signify a bad model. Another potential limitation of (static) DR plots is that they may offer little incremental value beyond confirming existing hypotheses:

\begin{quote}
``I would say I don't use tSNE that much, because it's hard for me to pose different questions to it... If it's already in a very beautiful space, like if it's already learned what I wanted it to learn, then tSNE is very useful.... what it's showing you is interpretable, because you've already interpreted it before.... But other than that — [if] you've learned an embedding, and you don't know what it does... then tSNE is not that useful. Because the clusters usually tend to be mixed. If you have an embedding that doesn't do well, let's say, then the projections are all mixtures of things... What is the axis that it's combining on? Maybe you don't know that. And it doesn't really give you those answers.'' (P6)
\end{quote}
Other participants echoed a similar sentiment relating to datasets with large numbers of unlabeled points, where the lack of known structure leads to a ``blob of points'' in the visualization (P2, P7).

Participants' doubts about DR may be surprising, particularly given that several techniques have been developed to visualize distortions and errors in DR \cite{Seifert2010,Stahnke2016,Nonato2019}. This consensus may have arisen because participants were not experts in DR, although a few participants had occasionally used visual DR tools such as Embedding Projector \cite{Smilkov2016} and Sleepwalker \cite{Ovchinnikova2019} (P6, P7, P8). Overall, participants perceived limitations in the fixed, error-prone distance transformations induced by DR, which led them to rely on their own handmade code snippets to probe embeddings.

\subsubsection{Nearest neighbors.} One of the most common methods that participants used to analyze embeddings, especially in the natural-language domain, was to identify points of interest and examine their nearest neighbors in the high-dimensional space (P1, P2, P3, P6, P8). When asked about the importance they placed in understanding embeddings at an individual-point level, participants gave a wide variety of responses depending on their roles and intended use cases. Those who were explicitly performing qualitative analyses on embeddings (P1, P2) used nearest neighbors extensively. For some participants who were building and validating models for downstream use, nearest neighbors were seen as an essential tool for exploration and debugging, even ``the most important tool we have'' (P5, P6). However, others who work with more heavyweight models and larger datasets (especially in industry) saw it as too time-consuming except to produce concrete demonstrations of results (P4, P9). In other words, while nearest neighbors serve as a very useful indicator of quality, it may currently be too difficult without \textit{a priori} points of interest to find samples that can be subjectively incorporated into a larger analysis.

\subsubsection{Clustering and feature analysis.} In addition to looking at individual points, some participants also considered groups of points as units of analysis, often naming standard techniques such as $k$-means and hierarchical clustering (P5, P8). For example, P9 described characterizing an image embedding model's weaknesses by probing the defining features of its clusters:

\begin{quote}
``We were seeing that things were clustering based on surface-level features, and we wanted it to be... clustering based on more semantic features. So we were looking at clusters as a way to understand what notion of similarity the feature representation is capturing, and whether that's the notion of similarity that we actually want to capture.'' (P9)
\end{quote}
In addition to understanding why points were grouped into a single cluster, participants were also interested in explaining why multiple clusters were close to each other; for example, in computational biology, cell types that cluster close together in an embedding space could reveal an underlying biological similarity (P5, P7). Overall, participants were interested in using clustering techniques, but tended to rely on simple programmatic tools to do so.

\subsubsection{Other techniques.} Two participants mentioned using axis-based analysis to understand embedding model behavior, i.e. studying the characteristics of points embedded along an axis between two words in the space, such as ``man'' versus ``woman'' (P2, P6). Similarly, P6 described probing generative image models by interpolating along an axis in the feature space between two images. Participants also described attempting or wanting to computationally assess the embedding topology, for example by characterizing the smoothness, continuity, or density of the space (P1, P6, P9). For simplicity and to adopt well-established techniques, we focus on the three main techniques described above as our design focus.

\subsection{Needs and makeshift strategies for embedding comparison}\label{sec:interview-comparison}

Unlike the techniques that participants use to understand individual embedding spaces, approaches for comparing more than one embedding space at a time were ``very rudimentary'' (P8) and few in number. Below, we discuss some common needs that experts expressed for embedding comparison, and the often-makeshift strategies they used to address them.

\subsubsection{Finding points that differ between spaces.} When asked about the types of comparative analyses they would like to conduct, one participant answered that they would simply like to ``identify the points that move the most'' (P5). Two of the NLP practitioners echoed a similar need (P1, P2); one described a strategy to sample pairs of points at different distances in one space, then calculate the distances between those points in another space. Similarly, while working on an image representation learning problem, P9 compared a new model against a baseline by examining point pairs with the smallest ratio of distances between the new embedding space and the baseline. Participants wanted to use this type of analysis both to gain insight into different datasets (P1, P2), and to qualitatively explain why a model works better or worse than another (P4, P9).

\subsubsection{Alignment and multimodal embeddings.} Because different embedding spaces usually have very different feature axes even if trained on the same data, a few participants mentioned that it would be helpful to be able to align embedding spaces, particularly for multimodal embeddings (e.g. joint embeddings of images and text, or texts from different languages). Embedding space alignment is an active area of research; nevertheless, participants did use some simple alignment techniques, such as aligning with respect to a single center point or performing a Procrustes alignment (minimizing root-mean-square distance between points) (P3, P5).

\subsubsection{Effects of modeling choices throughout the model-building pipeline.} Many model-builder participants expressed a desire to understand how model architectures and parameters were affecting the results at a more granular level than macroscopic accuracy metrics. For example, P7 was studying a novel dimensionality reduction technique, and wondered whether there was a better way to help users choose parameters than just looking at several visualizations side-by-side. P6 noted that when choosing a model architecture among many disparate options, they would generally compare the results manually before running a grid search over hyperparameters for the best architecture. Most participants' approaches to comparing model variants were limited to top-level accuracy numbers, which became insufficient in light of more subjective or instance-level requirements on the learned representations.

\subsection{Implications and design goals}\label{sec:design-goals}

Based on the techniques and needs that participants described to us, we generated four overarching themes and associated design goals to guide the development of our system:

\begin{enumerate}[label=\textbf{Goal \arabic*.}]
    \item \textbf{Facilitate greater model understanding by simplifying the process of comparing across multiple embedding spaces.} Participants described a fundamental desire to intuitively understand how a model is working and what notion of similarity an embedding space is capturing. However, characterizing embedding structure and similarity is challenging to do within a single embedding space. Some participants described validating embedding spaces by comparing examples to their prior knowledge about the specific data (P6, P7), while others develop a more long-term ``intuition around what's going on across different models and across different approaches'' (P8). We propose that supporting exploratory comparison between embedding spaces can help users understand \textit{individual} embedding spaces better by providing reference points from which to differentiate an embedding of interest.
    
    \item \textbf{Support exploration of large datasets by guiding the user to points and clusters that change meaningfully between embedding spaces.} Some participants enter the embedding analysis process equipped with specific points to analyze, but many do not extensively inspect the embeddings despite believing that it would be beneficial to do so. This discrepancy may arise because for many large datasets, the lack of labels to differentiate clusters makes it difficult to visually or programmatically pinpoint areas of interest. For example, after developing a new dimensionality reduction technique, P7 described the challenge of interpreting the results on a new dataset: ``We run this embedding, [and] we get this cloud of all blue points because we don't know how to color them. What do we do next?'' We propose to mitigate this complexity by developing recommendation features that guide the user to meaningful changes, which may also provide a form of clustering.
    
    \item \textbf{Support exploration of high-dimensional neighborhoods so users can avoid being misled by distortions due to DR projections.} As discussed above, participants were largely skeptical of the ability of DR to accurately capture the structure of an embedding space. In fact, many tended to avoid DR-based tools entirely in order to avoid drawing misinformed conclusions. However, participants did find DR helpful to get an initial impression of the embedding space, and to communicate their results. We hypothesize that when complemented by appropriate comparison tools, DR plots can serve as an effective ``map'' of the data that enables intuitive navigation and exploration.
    
    \item \textbf{Support integration into custom embedding analysis workflows.} The diversity of techniques we observed in the interviews indicates that embedding analysis and comparison often require custom, task-specific routines. For example, experts tend to have predefined hypotheses about what characteristics of points to search for, and they utilize specific downstream analyses to assess the embeddings of those points. An effective analysis tool will need to not only provide prebuilt methods for exploration and analysis, but allow them to move between the system's and their own analysis routines.

\end{enumerate}

\section{System Design}

We now introduce Emblaze, a system we developed based on the above design goals that seeks to help model builders compare notions of similarity and reliability in embedding spaces. Although Emblaze can be run as a standalone application, it is primarily designed as a widget that can be displayed in an interactive notebook environment. The tool integrates most of the major techniques described in the interviews, including nearest-neighbor analysis, clustering, and embedding space alignment. In fact, nearest neighbors form the backbone of most of the algorithms embedded in Emblaze, echoing previous embedding comparison efforts \cite{Heimerl2020} and reflecting the importance of nearest neighbors to interviewees across domains. Below, we provide an overview of Emblaze's interface followed by the features we developed to support each of the four design goals.

As depicted in Fig. \ref{fig:teaser}, Emblaze centers a dimensionality-reduction plot of the dataset that facilitates navigation and selection of points of interest. DR maps can be generated using common techniques (PCA, tSNE, and UMAP) along with a variety of distance metrics (cosine distance is the most common). To control for visual differences due to DR, Emblaze allows users to generate projections using AlignedUMAP, a variant of UMAP that adds a similarity constraint to the objective function \cite{McInnes2021}. Additionally, the projections are optimally scaled and rotated using Procrustes alignment to minimize coordinate differences \cite{Nguyen2019}. To the left of the main scatterplot is a panel listing the embedding spaces being compared, which we term ``frames''; clicking once on a thumbnail opens the comparison interface, and clicking again animates the points in the scatterplot smoothly to their locations in the new embedding space. Meanwhile, the right-hand sidebar contains a variety of tools to manage and analyze selections in the interface, including the nearest neighbors of the current selection, a browser for saved and recent selections, and Suggested Selections.

Neighbor-based metrics are the primary way to compare embedding vectors in Emblaze, because they can be computed in the high-dimensional space and are compatible with any quantitative distance metric. For a point $x$, we define $N_E(x)$ as the set of $k$ nearest neighbors to $x$ in the embedding space $E$ ($k$ is a constant that can be configured by the user, but is set to 100 by default). The rank of a neighbor $y$ in the neighbor set of $x$ is denoted $\text{rank}_E(y;x)$ and ranges from $0$ (first neighbor) to $k-1$. We frequently employ the Jaccard distance, denoted $J(\cdot,\cdot)$, to compare neighbor sets.

\subsection{Goal 1: Interactive embedding comparison views}\label{sec:frame-comparison}

Emblaze's primary function, addressing Goal 1, is the ability to compare embedding spaces using a combination of animation and explicit encodings. The tool's point selection features complement the animations by helping the user control the scope of the visualization, facilitating comparison at both instance-level and global scales. Screenshots of the various comparison views are shown in Fig. \ref{fig:diff_views}.

\subsubsection{Animation and Star Trails} 
\begin{figure*}
    \centering
    \includegraphics[width=\textwidth]{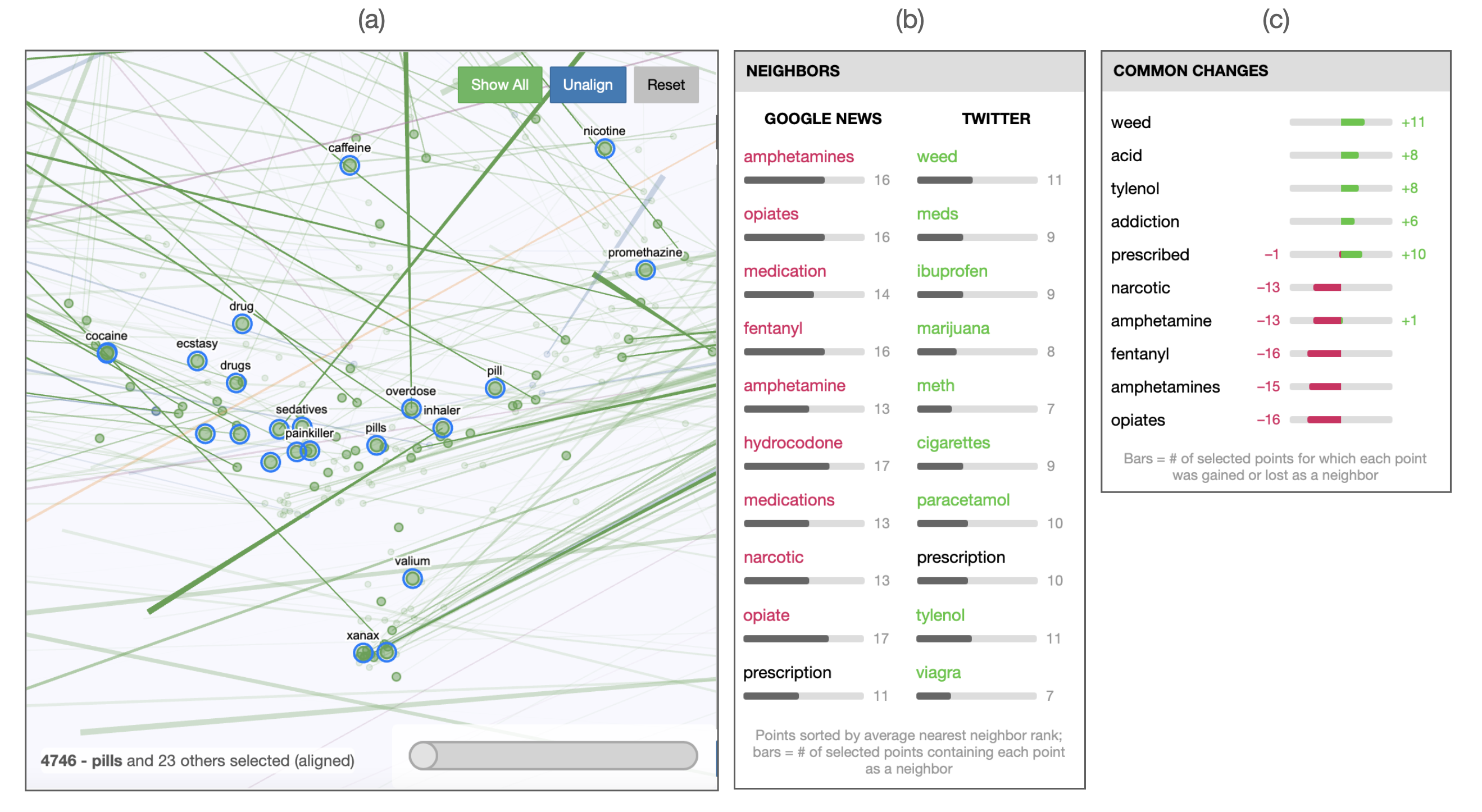}
    \caption{Comparison views for Google News vs. Twitter as in Fig. \ref{fig:teaser}, with a selection of drug-related words clustered by our suggestion algorithm. (a) The Star Trail visualization shows lines from each point's location in the Google News to its location in Twitter; line opacity indicates the proportion of nearest neighbors that have changed between the two spaces. Here, the diverging lines indicate that this neighborhood is breaking apart, creating clusters for illegal drugs, prescription pain medications, and others. (b) The neighbor differences view highlights the overall nearest neighbors of the selection in each space. (c) Common Changes indicate which neighbors are most commonly \textit{added} and \textit{removed} for this selection in Twitter (green and magenta, respectively). This highlights the addition of ``weed'' and ``acid'' as colloquial drug-related words in Twitter, and the removal of drug category names more characteristic of journalistic writing (``narcotic'', ``amphetamines'').} 
    \label{fig:diff_views}
\end{figure*}

When designing the animation between frames, we drew inspiration from the well-known Gapminder tool and other animated visualizations \cite{Heer2007}, which tap into the perceptual system's ability to track objects and identify motion outliers. However, some studies have found that animation introduces perceptual inaccuracy compared to small multiples and overlays \cite{Robertson2008}, and that motion outliers may be difficult to reliably perceive \cite{Veras2019}. These concerns may be exacerbated in DR plots, which often contain tens of thousands of points (orders of magnitude more than the scatter plots tested in the aforementioned studies). Therefore, we introduce an augmentation that we call a \textbf{Star Trail}\footnote{A star trail in astrophotography is a long-exposure image that shows ``trails'' of stars as they move through the sky due to Earth's rotation.}: a series of widening lines that show the paths of points from a source frame $A$ to a destination frame $B$, as shown in Fig. \ref{fig:diff_views}a. To draw the user's eye to larger changes, the opacity and width of each point $x$'s trail is set according to the proportion of nearest neighbors that have changed between the two frames, $1 - |N_A(x) \cap N_B(x)|/k$. Finally, a slider in the comparison view controls the animation progress from frame $A$ to $B$, allowing the user to drive the motion of the points along their respective Star Trail paths.

\subsubsection{Align and Isolate} The scatter plot's behavior in the comparison view can be adjusted based on the current selection, enabling the user to manage the visualization complexity and make many distinct comparative analyses between the same pair of frames. First, similar to the TensorFlow Embedding Projector \cite{Smilkov2016}, the user can Isolate to the current selection $S$ to limit the visible points to the selection and its immediate vicinity in every frame. Second, the visualization can be Aligned to the selected points' coordinates in the current frame, which recomputes the Procrustes transformation of every other frame such that the deviation of only the selected points is minimized. The combination of Aligning and Isolating to a selection effectively creates a smaller space for comparison and mitigates uninformative motion while animating the scatter plot. 

\subsubsection{Neighborhood differences and Common Changes} \label{sec:common-changes} The sidebar, which normally displays neighborhood information for the current selection, also shows comparison-specific information while the user is comparing two frames $A$ and $B$. First, a simple neighbor differences view shows the nearest neighbors of the selection in each frame (Fig. \ref{fig:diff_views}b); points are highlighted in magenta if they are present in $A$ but not $B$, and green if present in $B$ but not $A$. When multiple points are selected, an additional table of \textbf{Common Changes} lists the neighbors that are most commonly added to or removed from the nearest-neighbor sets of the selection $S$ (Fig. \ref{fig:diff_views}c). To compute Common Changes, we define a function to measure the inverse rank of the neighbors of $S$ that are present in one embedding but not another:
\begin{align}
    \text{Change}(y;A, B) = \sum_{x \in S}
\begin{cases}
    k - \text{rank}_{B}(y;x),& \text{if } y \cnot\in N_{A}(x) \text{ and } y \in N_{B}(x) \\
    0,              & \text{otherwise}
\end{cases} \label{eqn:common_changes}
\end{align}
Given this score function, the Common Change points can be computed as the top 5 and bottom 5 points in the union of all $S$'s neighbors $\bigcup_{x \in S}\left( N_A(x) \cup N_B(x) \right)$, scored using the criterion value $\text{Change}(y;A, B) - \text{Change}(y;B, A)$. When the number of nearest neighbors $k$ is sufficiently large (typically 50-100), these Common Changes can reveal change patterns that shed light on how frames $A$ and $B$ embed the selected points differently.

\subsection{Goal 2: Finding interesting comparisons}\label{sec:interesting-comparisons}

While the features described above help manage complexity and facilitate comparison given a previously-defined selection, we also sought to help the user find selections that yield interesting comparisons. For example, the Star Trail visualization preferentially highlights points with large nearest-neighbor differences that are also in the vicinity of the current selection, leading them to potentially interesting similar selections. We also developed a visual summary of change across frames, and an adaptive technique for suggesting selections, which we discuss below.

\subsubsection{Summary Color Stripes} We hypothesized that having a fast visual way to assess the amount of change in a group of points across all frames would accelerate the discovery of selections worth focusing on. Therefore, we developed a \textbf{Color Stripe} visualization that uses perceptual color similarity to encode similarity between frames for a given selection $S$. Color Stripes are determined by clustering the frames using a distance metric that captures both how much $S$ changes as a group with respect to its external neighborhood, and how much the neighborhoods \textit{within} $S$ change. More formally, the distance between frames $A$ and $B$ is computed as
\begin{align}
    d_{\text{frames}}(A,B;S) = \frac{1}{2|S|}\left(\sum_{x \in S} \Delta_{\text{inner}}(x;A, B, S) + \Delta_{\text{outer}}(x;A, B, S) \right) \label{eqn:color_stripe_distance}
\end{align}
where $\Delta_\text{inner}$ and $\Delta_\text{outer}$ reflect the change in the ``inner'' and ``outer'' neighbor sets of $x$ with respect to $S$:
\begin{align}
    \Delta_{\text{inner}}(x;A,B,S) &= \frac{1}{1 + |N_A(x) \cap N_B(x) \cap S|} \label{eqn:delta_inner} \\
    \Delta_{\text{outer}}(x;A,B,S) &= \frac{1}{1 + |N_A(x) \cap N_B(x) \setminus S|}  \label{eqn:delta_inner}
\end{align}

To visualize the clustering of frames generated by this distance metric, we assign each frame to a color along a ring in the CIELAB color space (a system in which Euclidean distance approximates perceptual distance). While the relative distances between frames around the ring correspond to differences in hue, the saturation of the colors is determined by the maximum distance between any of the frames. This results in highly consistent selections being represented as indistinguishable grayish hues, while highly varying selections feature bright colors. Examples of the Color Stripes can be seen next to the frame thumbnails to the left of the scatter plot (see Fig. \ref{fig:teaser}), as well as in the Suggested Selections pane (Fig. \ref{fig:suggested_selections}, right).

\subsubsection{Suggested Selections} 

Finding groups of points that exhibit meaningful, consistent neighborhood changes in large embedding spaces is challenging, particularly for groups of points that are not tightly clustered or labeled in the DR projection. Tightly clustered groups pose an additional problem: if a group shifts drastically between two spaces while remaining closely interconnected, the nearest neighbors of \textit{each point individually} may be largely similar even though the neighborhood \textit{around the cluster} has changed. To help users identify and navigate to such groups quickly, we created the \textbf{Suggested Selections} feature, which can be accessed through one of the sidebar tabs.

\begin{figure*}
    \centering
    \includegraphics[width=0.9\textwidth]{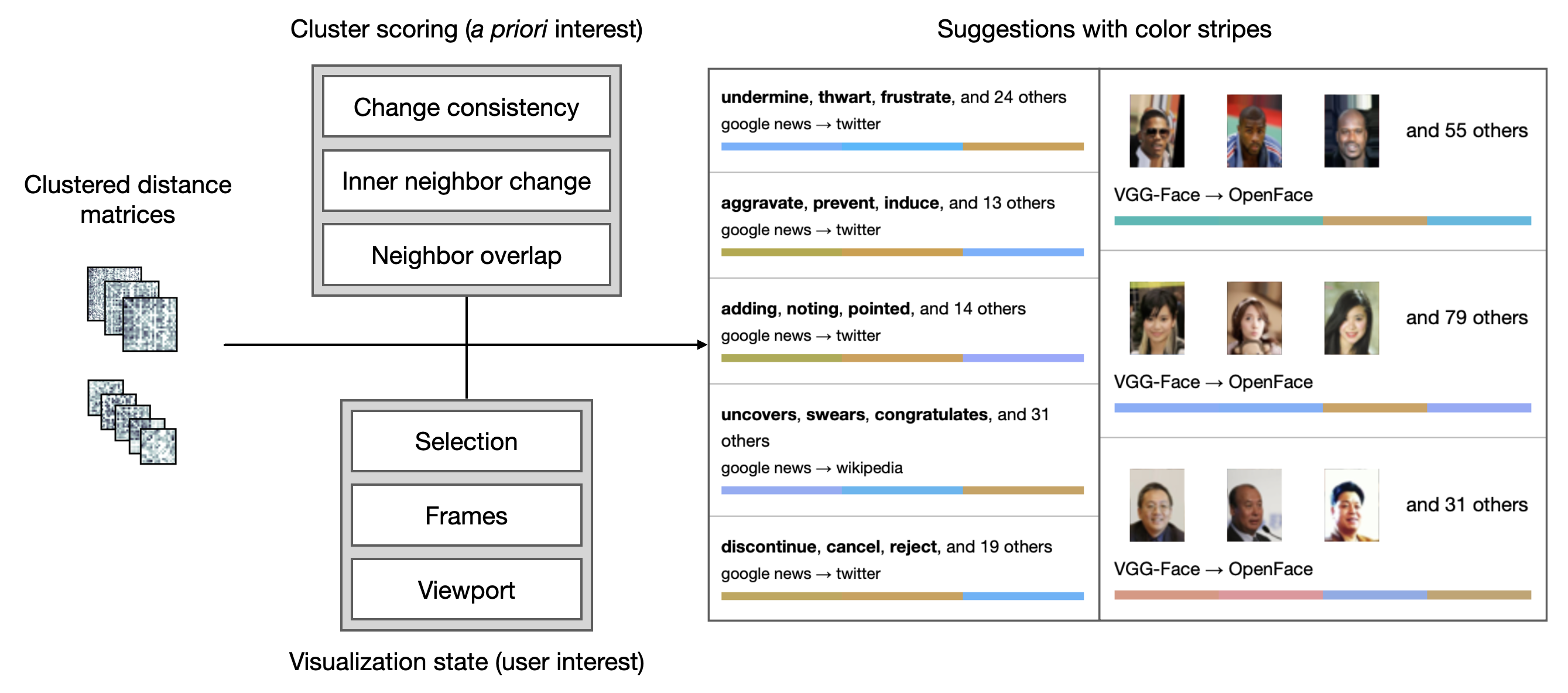}
    \caption{Process for generating Suggested Selections. First, pairwise point distances are precomputed to measure consistency of nearest-neighbor changes, and hierarchically clustered. Three cluster score metrics are used along with the current visualization state to filter and rank the clusters by their estimated degree of interest to the user. Finally, suggestions are displayed in the interface along with a Color Stripe to inform the user which frames have the most substantial variation. The example suggestions at right are drawn from the word embeddings comparison in Fig. \ref{fig:teaser} as well as a comparison of face recognition models on the CelebA celebrity faces dataset \cite{liu2015faceattributes}.}
    \label{fig:suggested_selections}
\end{figure*}

As shown in Fig. \ref{fig:suggested_selections}, the suggestion algorithm proceeds in two steps: one to precompute clusters, and one to rank and filter those clusters according to an interest function. In the precomputation step, a clustering of points is generated for each pair of frames $A$ and $B$ using a distance metric that measures the changes in neighbors gained and lost in $B$ compared to $A$:
\begin{align}
    d_{\text{points}}(x,y;A,B) ={} \frac{1}{2}( & J(N_B(x) \setminus N_A(x), N_B(y) \setminus N_A(y)) + {} \nonumber\\
    & J(N_A (x) \setminus N_B(x), N_A(y) \setminus N_B(y)) ) \label{eqn:suggested_selection_distance}
\end{align}
The effect of this formulation is that pairs of points which gain (or lose) a similar set of neighbors from frame $A$ to $B$ will have a small distance. The points are clustered using hierarchical clustering, with a variety of distance cutoffs to produce suggestion results of varying sizes. It is important to note that this clustering is quite different from a clustering performed within an embedding space, as is typically used \cite{LYi2015,Arendt2020}. Instead, the distance metric in Eqn. \ref{eqn:suggested_selection_distance} explicitly clusters points based on how they change from one frame to another, thereby directly codifying the types of change that we consider most ``interesting.''

In the suggestion phase, the clusters are ranked both by measures of \textit{a priori} interest and relevance to the current visualization state, resembling a classic degree-of-interest function \cite{VanHam2009}. The \textit{a priori} interest function is the sum of three metrics: consistency of neighbor gains and losses, changes in the cluster's inner neighbor structure (see Eqn. \ref{eqn:delta_inner}), and amount of neighbor overlap within the cluster. The clusters are then further filtered and ranked based on which frame(s) are being viewed, the currently-selected points and their neighbors, as well as the current bounds of the viewport. This enables the user to pan and zoom around the scatterplot and see Suggested Selections for each area they visit. 

Note that we have now proposed two distinct distance functions for seemingly similar purposes, namely $d_\text{frames}$ (Eqn. \ref{eqn:color_stripe_distance}) and $d_\text{points}$ (Eqn. \ref{eqn:suggested_selection_distance}). While the distance metric for Suggested Selections, $d_\text{points}$, helps to group together \textit{points} within a fixed pair of \textit{frames}, the metric for Color Stripes, $d_\text{frames}$, helps to group together \textit{frames} with respect to a fixed set of \textit{points}. This relationship is mirrored in how each clustering is used in the interface: users can select clusters surfaced by the Suggested Selections for a particular pair of frames, then use the Color Stripes to get a sense of their variations across all frames. By splitting the task of finding interesting comparisons into two complementary interactions, Emblaze extends the notions of interest established in prior work \cite{Heimerl2020,Boggust2019} to support both groups of points and more than two embedding spaces.

\subsection{Goal 3: Browsing high-dimensional neighborhoods}\label{sec:high-dimensional-space}

Since Emblaze is subject to the caveats of dimensionality reduction expressed in our formative interviews, we took care to distinguish the projection from the original high-dimensional space through visual augmentations and selection operations. These affordances are intended to encourage the use of the DR projection as a navigation tool by which users can find subspaces of interest, as described in Goal 3.

\subsubsection{High-dimensional neighbors}

When a point is hovered upon or selected, lines radiate outward from the point to its nearest neighbors in the high-dimensional space (by the distance metric pre-configured by the user). By assessing how far the lines extend while panning over the plot, the user can quickly see the fidelity of the 2D projection and find points that are far from their nearest neighbors.

The sidebar's neighbor list view, which has been established in prior embedding analysis systems \cite{Smilkov2016,Boggust2019,Ovchinnikova2019}, also supports showing neighborhoods for multiple-point selections. A selection $S$'s nearest neighbors in frame $A$ are the top 10 points in the full set of neighbors $\bigcup_{x \in S} \left( N_A(x) \right)$ that attain the best total inverse neighbor ranks with respect to the points in $S$. Note that simply listing the neighbors can be misleading if points have very disparate neighbor sets. Therefore, we also display a bar next to each neighbor indicating how many points in $S$ have that point as a neighbor; high values indicate a consistent neighborhood.

\subsubsection{High-dimensional radius select}

Emblaze provides a lasso selection tool to select points that are close together in 2D, which works well when a neighborhood is tightly clustered in the DR projection. For neighborhoods that are not well preserved by the projection, however, the user can use the Radius Select feature to find points within a configurable distance of a center point in the high-dimensional space.

\subsection{Goal 4: Computational notebook integration}\label{sec:notebook-architecture}

Towards the fourth goal, integrating embedding comparison into custom data science workflows, we implemented Emblaze as a widget that can be run in a Jupyter environment (as shown in Fig. \ref{fig:teaser}). This leads to unique benefits for users as compared to standalone applications. First, importing embedding data into Emblaze is made extremely simple, requiring only a set of coordinate arrays and images and/or text to describe each point. Second, once the viewer widget is instantiated within a Jupyter cell, the user can interact with the state of the system by manipulating either the interface or the underlying Python objects. For example, the \texttt{emblaze.Viewer} object exposes bidirectional properties for the current visible frame, comparison frame, selection, filter, and display settings. This enables interactions in which the user can visually identify a selection of interest, computationally analyze that selection using custom functions, identify new points of interest, then instantly navigate to the new selection in the interface.

\subsection{Availability and implementation details}

Emblaze is open-source and available on PyPI and GitHub\footnote{\url{https://github.com/cmudig/emblaze}}. The system consists of a Python backend, which performs most of the computationally-intensive analyses, and a Svelte frontend\footnote{\url{https://svelte.dev}}, which enables reactivity. The scatter plot is implemented using PIXI.js\footnote{\url{https://pixijs.com}}, a popular WebGL-based graphics framework. By implementing most plot rendering in custom shaders, Emblaze is able to display and animate tens of thousands of points smoothly on typical hardware.

\section{Case Studies}

As embedding comparison is a relatively nascent task in the literature on ML model analysis, we conducted case studies with ML experts to gain a preliminary understanding of how they might use Emblaze on real-world datasets. We recruited three ML expert researchers who were experienced in, and currently working with, embedding models or dimensionality reduction. The three users (whom we denote U1-3) prepared datasets from their work, installed Emblaze in their own programming environment, then engaged in a loosely scaffolded think-aloud analysis of their dataset. Participants spent 2-2.5 hours working with the investigators, and were compensated 20 USD per hour. The resulting audio transcripts and usage logs were used to build sequences of actions, which revealed how participants were using each Emblaze feature as part of their analyses.

Note that because Emblaze simply requires a set of embedding matrices and object descriptions, it is not limited to visualizing the final outputs of an embedding model. As depicted in Table \ref{tab:compare_levels}, Emblaze also supports other tasks such as comparing across different DR techniques, layers of a neural network, training data subsets, or corpora (e.g. for distributional semantic analysis \cite{covidconcepts2021}). The three experts' use cases and workflows discussed below represent just a few examples of how Emblaze can be utilized in practice.

\subsection{U1: Dimensionality reduction parameter selection}

U1's research centers around developing improved dimensionality reduction techniques, which necessitates comparisons of new techniques against existing ones on well-studied sample datasets. Here, they analyzed projections with different DR settings on the UCI Wine dataset, which contains physicochemical information and quality ratings for around 4,900 wines \cite{Cortez2009}. U1 chose to compare four projections of this dataset by manipulating 2 variables: projection technique (standard versus a custom implementation of UMAP) and \texttt{num\_neighbors} parameter (15 and 50).

\begin{figure*}
    \centering
    \includegraphics[width=\textwidth]{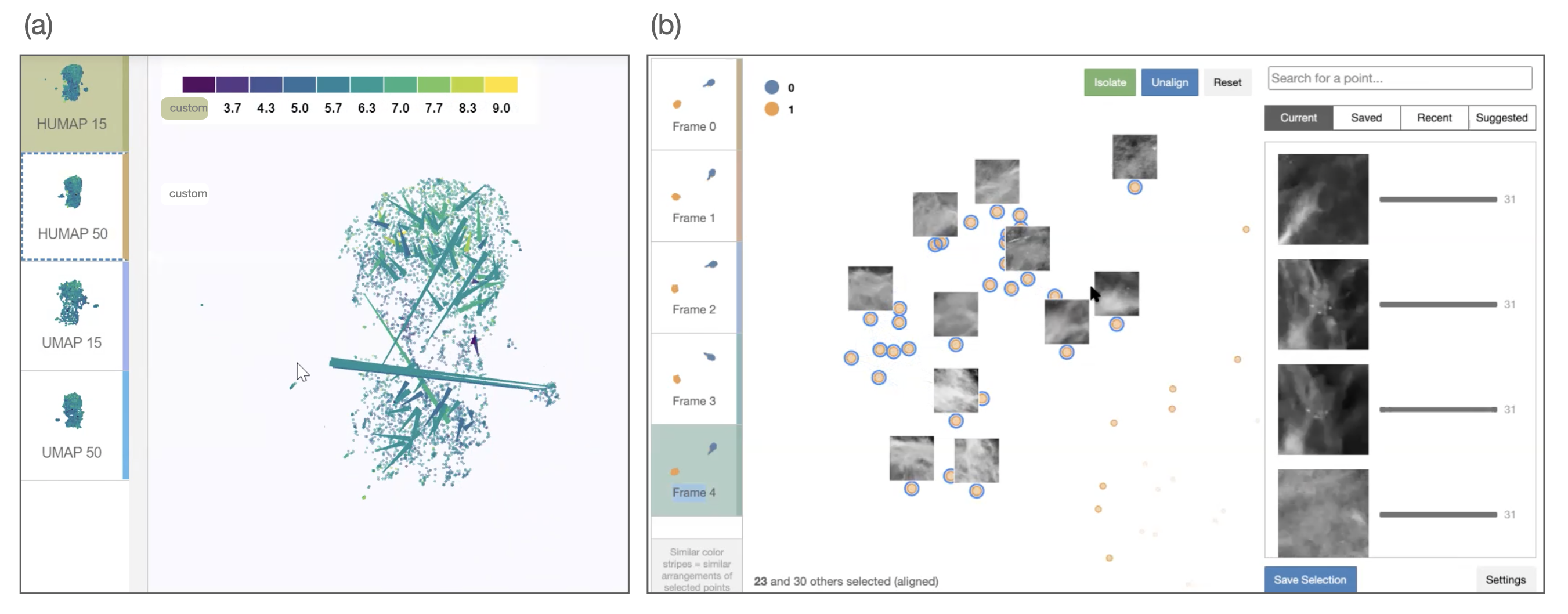}
    \caption{Screenshots of Emblaze showing (a) the UCI Wine dataset as explored by U1, and (b) a selection of mammogram patches investigated by U2.}
    \label{fig:u1-u2}
\end{figure*}

Using the Star Trail visualization to get an overview of changes between two projections of the custom UMAP implementation (shown in Fig. \ref{fig:u1-u2}a), U1 quickly observed that a large cluster of wines moved from one side of the plot to the other; they described this as an effect of UMAP's random initialization that they would like to overcome in their improved technique. They then used the manual animation slider to slowly interpolate back and forth between the two frames, and observed that the higher \texttt{num\_neighbors} parameter had resulted in a more compact, but less distinctly-clustered projection. To support this hypothesis, U1 performed a lasso selection and Isolated to a group of points while in the comparison view, causing the Star Trails to change to highlight changes relevant to that selection. Using the updated Star Trails and animation, they were able to watch the group move from being well-separated to integrated with the central mass of points.

Because the Wine dataset was mostly projected as one large mass of points with few well-separated clusters, U1 wanted to use the Suggested Selections feature to find differences within the central cluster. By scanning the Color Stripes in each suggestion, they identified a selection that differed considerably between the two \texttt{num\_neighbors} values. Looking at the sidebar, they perused the Common Changes for the selected group of points and noted that the most commonly added points were high-quality wines. Finally, they manually animated between the two frames while Aligned and Isolated to the selection. This revealed that the group of points was actually two clusters that were positioned next to each other when $\texttt{num\_neighbors} = 15$, but moved away from each other and acquired new neighbors when $\texttt{num\_neighbors} = 50$. This unexpected finding, which required looking at both Common Changes and geometric differences, highlighted tradeoffs between the two parameter choices that were not immediately visible before. After performing this analysis as well as a similar analysis on an unlabeled dataset of tweets, U1 reflected that comparing multiple projections was a much better initial step for exploring a new dataset: ``Especially [when] I don't have labels, I don't have a starting point in which way I can prepare my data set.... Using different embeddings... really helps you to get an overview according to various configurations of your data.''

\subsection{U2: Projection errors in medical image embeddings}

U2 is working on building machine learning models to distinguish breast cancer lesions from normal tissue in mammograms. At the time of the study, they were interested in analyzing the embedding space their model had learned in order to identify subtypes among the patches labeled as lesions. Therefore, they prepared a dataset consisting of 1,500 mammogram patches, each represented by a 2048-dimensional embedding vector. They then used UMAP to project the embeddings into 2D five times with random initializations; this would allow them to account for the effects of DR variation while exploring the embedding space.

U2's first step was to take an overview of the space by hovering over points to show their associated images, enabling them to characterize which regions corresponded to lesions and normal tissues. Then, using the Star Trails in the comparison view for two of the five DR variants, their attention was quickly drawn to a very long trail between the two main clusters, corresponding to a point labeled as a lesion that was projected with the normal patches in all but one frame. By looking at the points closest to the outlying point in each variant frame, they concluded that the point was likely a mislabeled normal patch that was in fact correctly embedded by the model. Using any one of the projections in isolation, this error case would likely have been missed.

U2 identified clusters of interest by selecting parts of the projection using the lasso-select tool, since they had recently performed a $k$-means clustering of the embedding space and found several contiguous regions that appeared to be meaningful. In one case, they selected a group of points that assumed two different geometries across the five frames: three frames were colored blue-green in the Color Stripes visualization, and the other two colored orange (shown in Fig. \ref{fig:u1-u2}b). U2 then opened the comparison view between one of the blue frames and one of the orange frames, and animated between the two to examine how their vicinities changed. By visually inspecting the points that the cluster moved towards, they hypothesized that the orange variants were less accurately isolating the selected neighborhood (although they noted that a medical expert would be needed to confirm which variants were more accurate). 

Looking at the variations between DR projections helped U2 gauge the reliability of projections, as well as the possibility of labeling errors: ``If some points move a lot, I would want to check them out, see if there's a problem with my data.'' Conversely, U2 was also excited that Emblaze allowed them to identify groups of points that were \textit{consistently} projected across different variants, indicating that those relationships were likely stable in the high-dimensional space. For example, they found a Suggested Selection whose patches all depicted marginal areas of the breast, and for which the Color Stripes were all gray (minimal variation between frames). Despite the fact that these points were not all mutual nearest neighbors in the projection, the constancy in their arrangement across multiple initializations provided a strong signal that the model considered them similar. Supporting Design Goal 3, U2 expressed that this assessment of consistency was ``definitely, definitely helpful, because there's no way for me to tell'' which parts of a projection are reliable otherwise.


\subsection{U3: Model comparison for knowledge graph representation learning}

\begin{figure*}
    \centering
    \includegraphics[width=\textwidth]{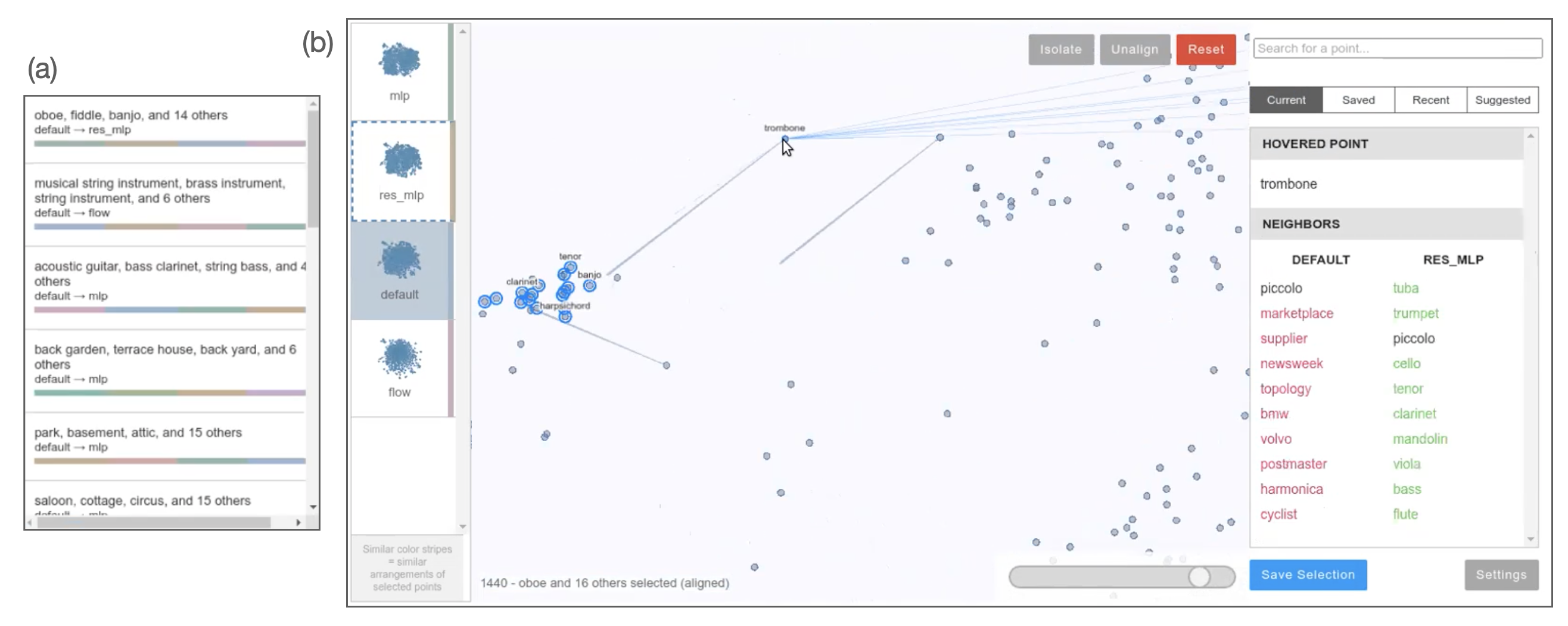}
    \caption{Suggested Selections for a region of U3's entity embedding model (a), and an examination of a point that joins one of the recommended clusters while animating between the default and supervised spaces (b).}
    \label{fig:u3}
\end{figure*}

U3 is a natural language processing expert working on building embedding representations of knowledge graphs (networks in which nodes represent entities and edges encode facts relating those entities). Starting from a pre-trained BERT model that simply encoded the text of each node, they had developed two versions that were fine-tuned to the facts in the knowledge graph (supervised), as well as a version that transformed the embedding space using a normalizing flow. They were aware that both models performed better than the base BERT model on a downstream task, but they lacked specific examples of how the embedding space structure had changed to yield the improved metrics. Therefore, they loaded a dataset of 5,000 sampled entities from a common-sense knowledge graph, embedded according to the four models. (Since the time of the session, Emblaze has been optimized to visualize many more points, mitigating the need for downsampling.) The four models were jointly projected into 2D using AlignedUMAP, which computes a UMAP with an additional loss term penalizing deviations between frames.

Initially, U3 focused on the comparison between the base and the two supervised models. They first selected a fairly well-separated cluster in the base model which consisted of color words. Then, opening the comparison view to look at the cluster's Common Changes between the default and supervised spaces, they found that the default model was including several phrases that matched the words in the cluster but not their semantic roles (e.g. ``blue umbrella,'' ``yellow ribbon''). Similarly, U3 lasso-selected and Aligned to a cluster of phrases in the base model that contained the word ``friend,'' then animated to the supervised model to see those points migrate apart from each other in the space. These examples confirmed U3's prior hypothesis that the base BERT model was overly reliant on lexical similarity compared to the fine-tuned version.

U3 was eager to use the Suggested Selections feature to find clusters of interest, in particular because the dataset had no labels that could serve as a color encoding for the scatter plot. For example, one suggestion they loaded was a set of 17 points comprising musical instruments (shown in Fig. \ref{fig:u3}a). In the comparison view between the base and supervised representations, they noticed that the Star Trail visualization was highlighting a few points moving into the cluster. They froze the transition between the two frames and navigated to the origin of each trail, revealing that the cluster was being augmented by less common instruments such as ``bagpipe'' and ``piccolo'' (Fig. \ref{fig:u3}b). They then selected these points individually to scan their neighbor differences views, and concluded that ``in the default space, it's just kind of garbage. But for the new space, it's a bunch of instruments. So that's actually very straightforward.'' Towards Design Goal 4, U3 voiced the importance of verifying these facts in the knowledge graph and did so directly in the notebook by extracting and looking up the selected point IDs.

Comparing the base model against the normalizing flow representation posed a more challenging task, because they expected the flow model to redistribute the space in a ``very noisy, not interpretable'' way. First, they animated between the two frames several times, noting that many of the clusters in the base model were less obvious in the flow variant. To make sure that these patterns were not just due to the projection, they browsed the Suggested Selections to find a cluster of interpretable points and eventually arrived at a cluster of concepts related to outer space. As above, they browsed the neighbor differences and Common Changes views to find that other space-related terms were commonly being removed in the flow model, while the neighbors most commonly added in the flow model were less sensible. Contemplating the differences between the supervised and flow models with respect to the base, U3 noted that
``the benefits [of flow] are not coming from improved alignment [of clusters], it's actually coming from the structure of the space... whereas [the supervised models] do seem to be helping for different reasons.'' The finding that these two variants both improved quantitative performance but in very different ways prompted U3 to think about new model architectures that could leverage the complementarity between the two methods.


\subsection{Overall user perspectives}

All three users thought the tool would be helpful in their work as (1) an interactive interface for DR projections, (2) a way to sanity-check their observations, and (3) a source of concrete examples to complement quantitative model performance metrics. Echoing the deep skepticism of DR expressed in the formative interviews, U3 noted that ``there are some inherent caveats with some of the reduction techniques,'' but that Emblaze, ``if anything, highlights and brings more attention to those'' through the high-dimensional neighbor comparison views. In support of Design Goal 2, all three users agreed that the Suggested Selections feature was very useful - particularly when there were no supervised labels to visually separate clusters. Overall, participants found that Emblaze made possible exploratory analyses they had never been able to do: ``I think there is a lot of functionality that I would never interface with if the tool didn't exist'' (U3).

Participants also provided useful feedback on the novel visual augmentations used in Emblaze. All three users were initially confused by the Color Stripe visualization, although they agreed that getting a sense of variation across all frames was important: ``Yeah, we definitely need that information, it's very helpful. I wish it was more straightforward'' (U2). Participants also wanted the Star Trail visualization to communicate more information about points' relationships to a cluster, such as by highlighting trails differently depending on whether they were entering or leaving a neighborhood (U1). Finally, participants thought notebook integration was very helpful for studying models without leaving their work environment, and suggested that the tool could integrate directly with models to dynamically compute and visualize embeddings for new groups of instances.





\section{Discussion}

By building upon designs from prior work as well as experts' current approaches to analysis and comparison, Emblaze enables a series of comparative workflows on embedding spaces that would have been highly challenging with existing tools. Our think-aloud sessions with ML experts suggest that the tool makes substantial progress towards the four goals described in Sec. \ref{sec:design-goals}, while revealing new possible directions for improvement:

\begin{enumerate}[label=\textbf{Goal \arabic*.}]
    \item \textbf{Facilitate greater model understanding by simplifying the process of comparing across multiple embedding spaces.} Both users who were working with learned representations (U2 and U3) gained new insights into the structure of their embedding spaces, using the Star Trail visualization, neighbor differences, and Common Changes views. 
    \item \textbf{Support exploration of large datasets by guiding the user to points and clusters that change meaningfully between embedding spaces.} All three users made extensive use of the Suggested Selections feature, particularly when clusters were not well separated by the projection, and found that it worked very well for their datasets. Participants had difficulty reading the Color Stripes visualization at first, a challenge that could be mitigated by simplifying the color encoding and giving it a dedicated space in the UI. However, they all agreed that Emblaze's ability to guide them to interesting and meaningful regions was a powerful addition to their workflow.
    \item \textbf{Support exploration of high-dimensional neighborhoods so users can avoid being misled by distortions due to DR projections.} Users agreed that animating between different DR projections and looking at the neighbor lists was a useful way to disambiguate between artifacts of the projection and true high-dimensional neighborhoods. It may be possible to assist the user's interpretation of these features to make them even more accessible to non-experts. For example, the interface could prompt the user to check the accuracy of a cluster when it is more disparate in the high-dimensional space than it appears in the projection.
    \item \textbf{Support integration into custom embedding analysis workflows.} Participants strongly favored Emblaze's notebook implementation over a standalone application, primarily because of ease of installation and compatibility with data that participants had previously stored. They also suggested new visualization possibilities if Emblaze were even more tightly integrated with ML frameworks in the future.
\end{enumerate}

The case studies presented here cannot be interpreted as a comprehensive evaluation of Emblaze's features, particularly since our users had not used similar tools before and had no baseline for comparison. Rather, our observations point to novel workflows that model builders can utilize through Emblaze and that can be built upon in future work. Echoing the needs expressed by our interview participants, many of these workflows led to a greater understanding of the \textit{notions of similarity} that embedding spaces were capturing. For instance, U3's use of Suggested Selections enabled them to quickly find several clusters that diverged from one model to another in similar ways. By identifying common patterns of change across these clusters, they were able to construct a narrative for how the architecture choices underlying each model had resulted in the differences they observed. This process was made much more efficient by U3's back-and-forth interaction between the visualization and code, not only to corroborate findings for large groups of points, but also to quickly load up multiple subsets of the data for a more robust analysis.

Emblaze also afforded our expert users new workflows that helped them reason about the \textit{reliability} of their embedding space analyses, and what conclusions they could sensibly draw from them. For example, even though U2 was only examining one model space, they were more confident in identifying reliable clusters because they could assess their consistency across DR projections. Comparison also allowed users to easily assess the quality of embedding neighborhoods, which would ordinarily require an intuition built up over many past experiences analyzing embedding spaces. After finding a cluster with substantial variation between two models, for instance, U3 could easily conclude that the cluster was poorly embedded in one model because its neighbors made little sense \textit{relative to} the more reasonable neighbors in the other model. With enough experience, it would likely be possible to draw similar conclusions based on a single embedding space; Emblaze has the potential to help users build these intuitions more quickly.

Some features that would be important for particular use cases were omitted from this first version of Emblaze for simplicity. Most notably, a few interview participants described wanting to know what feature axes drive the separation of a cluster, e.g. which genes are highly expressed in a particular cluster of cells from a computational biology experiment. Although Emblaze's support for text and image data types covers many typical ML representations, incorporating visualizations to help users interpret points and clusters in tabular data (such as those proposed in prior embedding analysis tools \cite{Stahnke2016}) could expand the tool's applicability even further. In addition, Emblaze has only one scatter plot view that animates between projections, a promising alternative to prior work that juxtaposes multiple projections next to each other \cite{Boggust2019,Arendt2020,Cutura2020}. In the future, though, the two approaches could be combined by allowing users to toggle between a single animated scatter plot for large-scale browsing and side-by-side visualizations to compare smaller subsets of the data.

\section{Conclusion}

In this work, we have synthesized experts' viewpoints across different domains to construct a tool that enables visualization and exploration across several embedding spaces--previously an extremely difficult task. Our limited evaluation suggests that the system considerably lowers the barrier to embedding analysis and comparison. However, further engagement with model builders as well as non-expert users (such as ML students) is needed to determine how visualization tools can support these tasks even more effectively. Given the increasing societal impact of ML models trained on vast unlabeled datasets, qualitative comparison may help track our progress towards more valid, unbiased, and ethical representations. By making Emblaze open source and publicly available, we hope to spark experimentation and discussion in the ML and visualization communities on how embedding space comparison can help produce more accurate and responsible models.


\begin{acks}
This work was supported by the Center for Machine Learning and Health at Carnegie Mellon University. We thank Carolyn Rosé, Jill Lehman, Denis Newman-Griffis, and Dominik Moritz for their valuable feedback throughout the development of Emblaze. Thanks also to Ángel (Alex) Cabrera for laying the groundwork for reactive Jupyter widgets, which made the notebook implementation of Emblaze possible. Finally, we are grateful to all our study participants for sharing their time and insights.
\end{acks}

\bibliographystyle{ACM-Reference-Format}
\bibliography{references}










\end{document}